\documentclass[prb,superscriptaddress,showpacs,twocolumn]{revtex4}

\usepackage{epsfig}

\begin{document}

\title{Integrated complementary graphene inverter}

\author{Floriano Traversi}
\affiliation{L-NESS, Department of Physics, Politecnico di Milano, Polo
Regionale di Como, Via Anzani~42, 22100~Como, Italy}
\author{Valeria Russo}
\affiliation{Micro and Nanostructured Materials Laboratory, Department of
Energy, Politecnico~di~Milano, Via Ponzio 34/3, 20133 Milano, Italy}
\author{Roman Sordan}
\email{roman.sordan@como.polimi.it}
\affiliation{L-NESS, Department of Physics,
Politecnico di Milano, Polo Regionale di Como, Via Anzani~42, 22100~Como,
Italy}

\date{\today}

\begin{abstract}
The operation of a digital logic inverter consisting of one $p$- and one
$n$-type graphene transistor integrated on the same sheet of monolayer graphene
is demonstrated. The type of one of the transistors was inverted by moving its
Dirac point to lower gate voltages via selective electrical annealing. Boolean
inversion is obtained by operating the transistors between their Dirac points.
The fabricated inverter represents an important step towards the development of
digital integrated circuits on graphene.

\end{abstract}


\maketitle

Graphene, a recently isolated\cite{novoselov04} single sheet of graphite, is
currently being investigated as a viable alternative to Si for the channel of
field-effect transistors~(FETs) at the \mbox{sub-10~nm} scale, at which the
ultimate limits of Si technology would probably be reached.\cite{meindl01} The
high mobility of carriers in graphene\cite{bolotin08,du08} could allow
fabrication of FETs with a very low channel resistance, resulting in a high
operational speed.\cite{lin09} The remarkable electronic properties of
graphene\cite{neto09} and its compatibility with Si lithographic
techniques\cite{han07,lemme07} promise to simplify the transition to
carbon-based electronics.\cite{avouris07} Large-scale fabrication of graphene,
which is currently being attempted by epitaxial growth,\cite{zhou07,reina09}
transfer printing,\cite{liangfu07,lidong08,liang09} or deposition from a
solution,\cite{tung08} is the following step in the development of
graphene-based integrated circuits. However, only single-transistor
operation\cite{geim07,li08,sordan09} has been demonstrated so far. Here we
demonstrate the operation of the first graphene integrated electronic circuit,
consisting of two graphene FETs of opposite type. The transistors are
fabricated on the same sheet of monolayer graphene and comprise an integrated
digital logic inverter (NOT gate), the main building block of Si complementary
metal-oxide semiconductor (CMOS) digital electronics.\cite{kang02}

\begin{figure}
\centering \psfig{figure=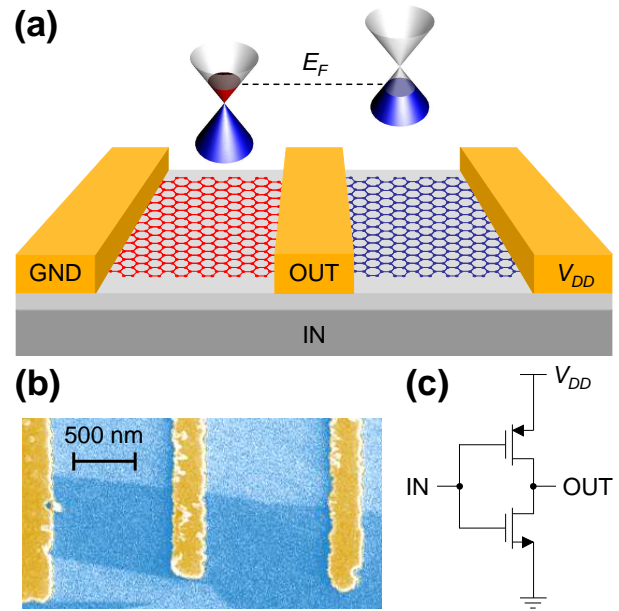, width=8cm} \caption{Integrated
complementary graphene inverter. (a)~A schematic of the fabricated inverter.
Three electrodes patterned on the same flake of monolayer graphene define two
FETs. The part of the flake between the two leftmost electrodes (depicted in
red) is electrically annealed to obtain an $n$-type FET. The other part of the
flake (depicted in blue) is a pristine $p$-type FET. The flake is electrically
insulated from the input (highly doped Si depicted in dark grey) by a layer of
SiO$_2$ (depicted in bright grey). (b)~Scanning electron microscopy image of
the fabricated inverter. Electrode separation (channel length) is 1~$\mu$m.
(c)~The circuit layout (power supply $V_{DD}=3.3$~V).} \label{fig:inverter}
\end{figure}

\begin{figure}
\centering \psfig{figure=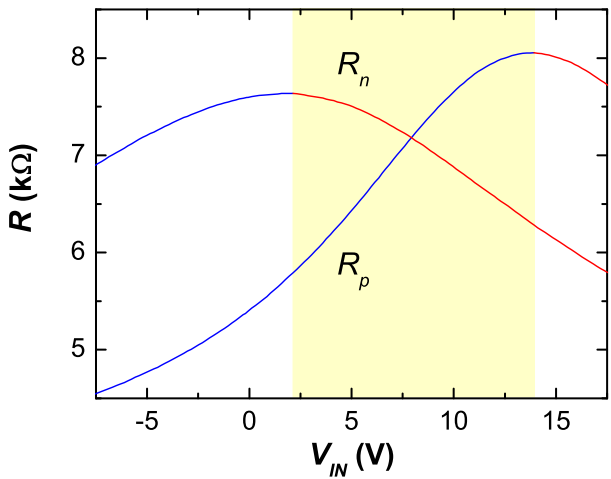, width=8cm} \caption{Resistance curves $R$
vs.\ $V_{IN}$ of the graphene transistors schematically depicted in
Fig.~\ref{fig:inverter} at $T=3$~K. The curve denoted by $R_n$ ($R_p$)
corresponds to the left (right) transistor. Segments of the curves displaying
$n$-type ($p$-type) behavior are drawn in red (blue). In the range (shaded in
yellow) between Dirac points (2.1~V~$<V_{IN}<$~13.9~V) transistors exhibit
complementary behavior, i.e., the Fermi level in the left (right) transistor is
in the conduction (valence) band, as depicted in Fig.~\ref{fig:inverter}(a).}
\label{fig:resistance}
\end{figure}

The fabricated inverter is schematically depicted in Fig.~\ref{fig:inverter}.
Graphene flakes were deposited by mechanical exfoliation of highly oriented
pyrolitic graphite on a highly doped Si substrate with 300~nm of thermally
grown dry SiO$_2$ on top.\cite{novoselov04} A metal contact evaporated on the
back of the Si substrate was used as a back-gate. The inverter was fabricated
on a flake which was identified as a monolayer graphene by Raman
spectroscopy.\cite{ferrari06} The flake was contacted by three
Cr(5~nm)/Au(50~nm) electrodes patterned by e-beam lithography. Each part of the
flake contacted by a pair of neighboring electrodes (source and drain contacts)
comprises a channel of one of the two graphene FETs, which share the same
back-gate used as voltage input (IN). Both FETs show identical $p$-type
behavior at small gate voltages, which has been attributed to hole-doping by
physisorbed ambient impurities such as water\cite{novoselov04} and
oxygen.\cite{liu08} The measured transfer resistance $R_p$ between the source
and drain contacts of one of the FETs [the right-hand one in
Fig.~\ref{fig:inverter}(a)] as a function of the applied back-gate voltage
$V_{IN}$ is shown in Fig.~\ref{fig:resistance}. The $p$-type behavior is
exhibited up to the Dirac point (resistance maximum) which is reached at
$V_{IN}=13.9$~V. For higher input voltages, the Fermi level crosses into the
conduction band resulting in type inversion.

In order to fabricate an inverter, one of the transistors [the left-hand one in
Fig.~\ref{fig:inverter}(a)] was electrically annealed\cite{moser07} to shift
its Dirac point to lower input voltages. Annealing was carried out in a He
atmosphere ($\sim 5$~mbar) at $T=3$~K and $V_{IN}=0$~V. The source-drain
voltage was increased from zero to 3~V in steps of 0.1~V to remove ambient
contamination by Joule heating. After each step, the voltage was held constant
for $\sim 5$~min. The highest drain current reached was $\sim 350$~$\mu$A. For
voltages larger than 2.5~V the drain current noticeably decreased during the
hold time, due to the removal of $p$-type impurities. After annealing, the
resistance curve of the annealed transistor $R_n$ vs.\ $V_{IN}$ was found to be
shifted to lower input voltages with the peak at $V_{IN}=2.1$~V, as shown in
Fig.~\ref{fig:resistance}. The annealing procedure did not affect the other
transistor, whose resistance curve $R_p$ vs.\ $V_{IN}$ (also shown in
Fig.~\ref{fig:resistance}) was unchanged. Hence, the transistors exhibit
complementary behavior (non-annealed as a $p$- and annealed as an $n$-type FET)
in the range 2.1~V~$<V_{IN}<$~13.9~V. Electrical annealing offers a simple way
to change the type of a selected transistor, in contrast with conventional
thermal annealing\cite{ishigami07} which affects all transistors on a chip.

\begin{figure}
\centering \psfig{figure=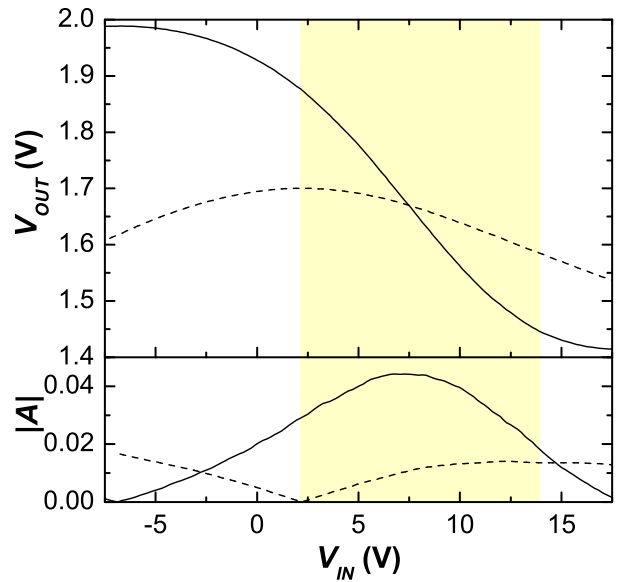, width=8cm} \caption{The measured DC voltage
transfer characteristics of the fabricated complementary graphene inverter
(solid lines) and a resistive-load inverter (dashed lines) obtained by
replacing the $p$-type transistor by a resistor. The characteristics are
represented by the output voltage $V_{OUT}$ and absolute value of the voltage
gain $A=dV_{OUT}/dV_{IN}$ as functions of the input voltage $V_{IN}$ at
$T=3$~K. The FETs exhibit complementary mode of operation in the range shaded
in yellow.} \label{fig:transfer}
\end{figure}

The inverter was realized by connecting the source of the $n$-graphene FET to
ground (GND), the source of the $p$-graphene FET to a conventional CMOS supply
voltage $V_{DD}=3.3$~V, and the output (OUT) to the common drain of the FETs
(Fig.~\ref{fig:inverter}).\cite{kang02} In this configuration, the output
voltage is given by $V_{OUT}=V_{DD}/(1+R_p/R_n)$. The measured voltage transfer
characteristics of the fabricated inverter are shown in
Fig.~\ref{fig:transfer}. The transfer curve $V_{OUT}$ vs.\ $V_{IN}$ can be
understood from the previous expression and Fig.~\ref{fig:resistance}, as the
graphene FETs stay in the ohmic regime even at very large drain biases. They
function as simple voltage controlled resistors whose resistances $R_p$ and
$R_n$ depend solely on the applied gate voltage $V_{IN}$. The two FETs operate
in the complementary mode between Dirac points; in this range, increase of
$V_{IN}$ causes resistance $R_p$ to increase and $R_n$ to decrease which
results in a strong increase in the ratio $R_p/R_n$. As a consequence,
$V_{OUT}$ decreases with the increase of $V_{IN}$ giving rise to the voltage
inversion shown in Fig.~\ref{fig:transfer}. Away from the Dirac points, the
output voltage saturates as both FETs enter the same mode of operation
($p$-type for $V_{IN}<2.1$~V and $n$-type for $V_{IN}>13.9$~V) making the ratio
$R_p/R_n$ approximately constant. However, in contrast with a CMOS inverter,
the output voltage does not saturate to zero or $V_{DD}$ as neither of the FETs
can be turned off. Inability to turn off the FETs stems from the absence of a
bandgap in graphene and the formation of electron-hole puddles.\cite{martin08}

The threshold voltage $V_{TH}$ of a logic gate is usually defined as the input
voltage at which the absolute value of the voltage gain $A=dV_{OUT}/dV_{IN}$
(shown in Fig.~\ref{fig:transfer}) reaches a maximum. This ensures the maximal
output voltage swing, i.e., a clear distinction between Boolean 0 and 1 at the
output. The maximum absolute gain of $|A|=0.044$ was reached at
$V_{IN}=V_{TH}=7.5$~V (Fig.~\ref{fig:transfer}). At this operating point $R_n$
is slightly larger than $R_p$ (the resistance curves in
Fig.~\ref{fig:resistance} intersect at $V_{IN}\simeq 8.0$~V), so the output
voltage is slightly larger than $V_{DD}/2$. The small voltage gain of the
fabricated inverter is due to a very small change of the resistance of the
transistors around the Dirac point, i.e., due to the impossibility of turning
the transistors off (the resistance off/on ratio in Fig.~\ref{fig:resistance}
is only $\simeq 1.8$). Although the small gain also suppresses noise, the logic
gates do not have a noise margin as the gain is always less than~1 and there is
a mismatch between the input offset $V_{TH}$ and output offset $\simeq
V_{DD}/2$  (in contrast with conventional CMOS gates where $V_{TH}=V_{DD}/2$).

Voltage inversion can also be obtained if one of the FETs is replaced by an
off-chip resistor.\cite{sordan09} However, in this case the corresponding
resistance in the ratio $R_p/R_n$ is constant so the output voltage $V_{OUT}$
decreases more slowly when $V_{IN}$ increases. To demonstrate this, an
$n$-graphene inverter was realized by replacing the $p$-type transistor with an
off-chip resistor of the same resistance at $V_{IN}=7.5$~V. The measured
voltage transfer characteristics of such a resistive-load inverter are shown in
Fig.~\ref{fig:transfer}. The much smaller output voltage swing and gain
obtained in this case stress the importance of the complementary mode of
operation.

Voltage transfer characteristics are presented at $T=3$~K in order to evaluate
the upper limit of performance of the fabricated inverter. The characteristics
were also measured at room temperature by keeping the inverter in vacuum ($\sim
10^{-2}$~mbar) so as not to reintroduce ambient contamination which would shift
the Dirac point of the annealed transistor back to the original
position.\cite{moser07} The principle of operation did not change at room
temperature, but broadening of resistance peaks degraded inverter performance.
The output voltage swing was damped and the highest measured absolute value of
the voltage gain was $|A|=0.027$. The supply voltage $V_{DD}=3.3$~V was found
to be too high for room-temperature operation as the electrical current was not
stable at $V_{IN}=0$~V (most likely due to Joule heating of the transistors).
At lower supply voltages ($V_{DD}=1.1$~V) the current was stable at a constant
input voltage, but the position of the peaks was found to slowly drift with
time, probably due to an inadvertent contamination of graphene by residual
gasses. In contrast, by keeping the sample at $T=3$~K the position of the peaks
did not change in 60 days, although the inverter was repeatedly measured with
the full supply voltage of $V_{DD}=3.3$~V.

\begin{figure}
\centering \psfig{figure=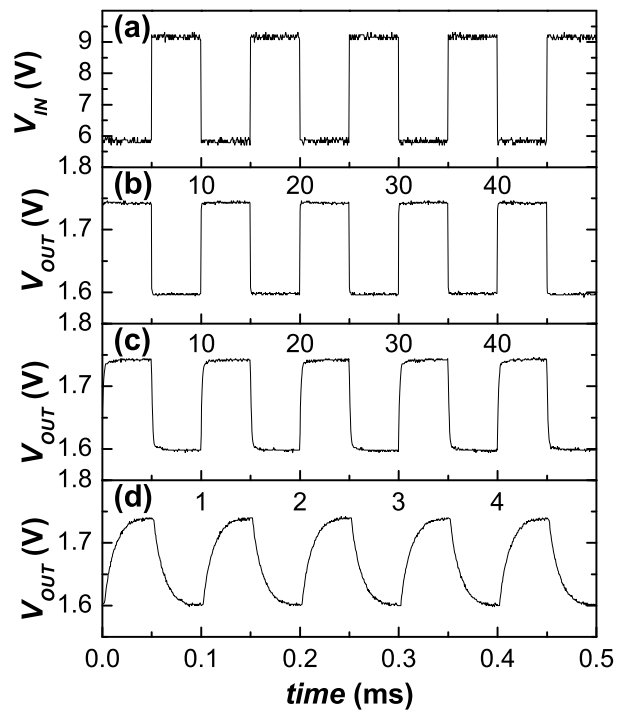, width=8cm} \caption{Digital waveforms
measured on the fabricated inverter. (a)~Input voltage. The offset is 7.5~V,
voltage swing $V_{DD}=3.3$~V, and frequency $f=100$~Hz. The following panels
show the output voltage at $T=3$~K with input signal frequency of
(b)~$f=100$~Hz, (c)~$f=1$~kHz, and (d)~$f=10$~kHz.} \label{fig:waveforms}
\end{figure}

Dynamic pulse response measurements of the fabricated inverter at three
different clock rates of the input signal are shown in
Fig.~\ref{fig:waveforms}. The measurements were performed by driving the
inverter with a square-wave signal with the offset $V_{TH}=7.5$~V. The total
input voltage swing was $V_{DD}=3.3$~V as in conventional CMOS logic gates.
Under these conditions, stable and separated output logic levels are obtained
at all frequencies, as shown in Fig.~\ref{fig:waveforms}. However, the output
voltage swing ($\simeq 0.15$~V) is much smaller than the input voltage swing
(3.3~V) because of the small voltage gain. As the clock rate of the input
signal is increased the output signal becomes more distorted and already at
10~kHz propagation delay can no longer be neglected. The large total parasitic
capacitance $C\simeq 3$~nF of the measurement equipment connected to the output
of the inverter and the output resistance of the inverter
$R=(R_p^{-1}+R_n^{-1})^{-1}\simeq 3.5$~k$\Omega$ limit the clock rate to
$f_{max}=1/(2\pi RC)\simeq 15$~kHz. In principle, by loading the output with a
typical gate capacitance of $C\sim$~10~fF,\cite{mistry07} a clock rate of
$f_{max}\sim 4.5$~GHz could be obtained. Further increase of $f_{max}$ by
reduction of length of the graphene FETs (to reduce $R$ by reducing the $R_p$
and $R_n$) will be hampered by unscalable contact resistance.

Although this inverter seems to be an attractive alternative to a Si CMOS
inverter, there are other two important figures of merit that should be
considered. First, the inverter is always conducting, i.e., the output stage
dissipates a static power $V_{DD}^2/(R_p+R_n)\sim 0.77$~mW, in contrast with a
CMOS inverter in which there is no static power dissipation. The static
dissipation could be reduced by using graphene transistors with a higher
resistance, but this would increase the transient response time making this
inverter much slower than a state-of-the-art CMOS inverter whose output
resistance is $\sim$~736~$\Omega$.\cite{mistry07} Hence, there is a trade-off
between the static dissipation and the highest possible clock rate. As in Si
CMOS technology, the clock rate will eventually be limited by power dissipation
rather than intrinsic transistor parameters.\cite{lin09} Second, input and
output logic voltage levels are not the same, so the inverters could not be
directly cascaded (i.e., level shifters would be required). This problem could
be mitigated to some extent by decreasing the input threshold $V_{TH}$ from the
present value of $7.5$~V to $\simeq V_{DD}/2=1.65$~V by annealing both
transistors. However, this would not increase the voltage gain so the mismatch
between the input and output voltage swing would prevent direct cascading of
the logic gates as long as gapless graphene is used.

In summary, a complementary logic inverter was fabricated by integrating two
transistors of the opposite type on the same flake of a monolayer graphene. The
voltage transfer characteristics of the fabricated inverter exhibit clear
voltage inversion. Dynamic pulse measurements display characteristic NOT
functionality when the inverter is operated with a CMOS input voltage swing and
supply voltage. Although application of the present inverter is limited by
power consumption and inability for direct cascading, its realization
demonstrates feasibility of using graphene as a substrate on which complete
electronic circuits can be integrated.


\begin{thebibliography}{25}
\expandafter\ifx\csname natexlab\endcsname\relax\def\natexlab#1{#1}\fi
\expandafter\ifx\csname bibnamefont\endcsname\relax
  \def\bibnamefont#1{#1}\fi
\expandafter\ifx\csname bibfnamefont\endcsname\relax
  \def\bibfnamefont#1{#1}\fi
\expandafter\ifx\csname citenamefont\endcsname\relax
  \def\citenamefont#1{#1}\fi
\expandafter\ifx\csname url\endcsname\relax
  \def\url#1{\texttt{#1}}\fi
\expandafter\ifx\csname urlprefix\endcsname\relax\def\urlprefix{URL }\fi
\providecommand{\bibinfo}[2]{#2} \providecommand{\eprint}[2][]{\url{#2}}

\bibitem[{\citenamefont{Novoselov et~al.}(2004)\citenamefont{Novoselov, Geim,
  Morozov, Jiang, Zhang, Dubonos, Grigorieva, and Firsov}}]{novoselov04}
\bibinfo{author}{\bibfnamefont{K.~S.} \bibnamefont{Novoselov}},
  \bibinfo{author}{\bibfnamefont{A.~K.} \bibnamefont{Geim}},
  \bibinfo{author}{\bibfnamefont{S.~V.} \bibnamefont{Morozov}},
  \bibinfo{author}{\bibfnamefont{D.}~\bibnamefont{Jiang}},
  \bibinfo{author}{\bibfnamefont{Y.}~\bibnamefont{Zhang}},
  \bibinfo{author}{\bibfnamefont{S.~V.} \bibnamefont{Dubonos}},
  \bibinfo{author}{\bibfnamefont{I.~V.} \bibnamefont{Grigorieva}},
  \bibnamefont{and} \bibinfo{author}{\bibfnamefont{A.~A.}
  \bibnamefont{Firsov}}, \bibinfo{journal}{Science}
  \textbf{\bibinfo{volume}{306}}, \bibinfo{pages}{666} (\bibinfo{year}{2004}).

\bibitem[{\citenamefont{Meindl et~al.}(2001)\citenamefont{Meindl, Chen, and
  Davis}}]{meindl01}
\bibinfo{author}{\bibfnamefont{J.~D.} \bibnamefont{Meindl}},
  \bibinfo{author}{\bibfnamefont{Q.}~\bibnamefont{Chen}}, \bibnamefont{and}
  \bibinfo{author}{\bibfnamefont{J.~A.} \bibnamefont{Davis}},
  \bibinfo{journal}{Science} \textbf{\bibinfo{volume}{293}},
  \bibinfo{pages}{2044} (\bibinfo{year}{2001}).

\bibitem[{\citenamefont{Bolotin et~al.}(2008)\citenamefont{Bolotin, Sikes,
  Jiang, Klima, Fudenberg, Hone, Kim, and Stormer}}]{bolotin08}
\bibinfo{author}{\bibfnamefont{K.~I.} \bibnamefont{Bolotin}},
  \bibinfo{author}{\bibfnamefont{K.~J.} \bibnamefont{Sikes}},
  \bibinfo{author}{\bibfnamefont{Z.}~\bibnamefont{Jiang}},
  \bibinfo{author}{\bibfnamefont{M.}~\bibnamefont{Klima}},
  \bibinfo{author}{\bibfnamefont{G.}~\bibnamefont{Fudenberg}},
  \bibinfo{author}{\bibfnamefont{J.}~\bibnamefont{Hone}},
  \bibinfo{author}{\bibfnamefont{P.}~\bibnamefont{Kim}}, \bibnamefont{and}
  \bibinfo{author}{\bibfnamefont{H.~L.} \bibnamefont{Stormer}},
  \bibinfo{journal}{Solid State Commun.} \textbf{\bibinfo{volume}{146}},
  \bibinfo{pages}{351} (\bibinfo{year}{2008}).

\bibitem[{\citenamefont{Du et~al.}(2008)\citenamefont{Du, Skachko, Barker, and
  Andrei}}]{du08}
\bibinfo{author}{\bibfnamefont{X.}~\bibnamefont{Du}},
  \bibinfo{author}{\bibfnamefont{I.}~\bibnamefont{Skachko}},
  \bibinfo{author}{\bibfnamefont{A.}~\bibnamefont{Barker}}, \bibnamefont{and}
  \bibinfo{author}{\bibfnamefont{E.~Y.} \bibnamefont{Andrei}},
  \bibinfo{journal}{Nature Nanotech.} \textbf{\bibinfo{volume}{3}},
  \bibinfo{pages}{491} (\bibinfo{year}{2008}).

\bibitem[{\citenamefont{Lin et~al.}(2009)\citenamefont{Lin, Jenkins,
  Valdes-Garcia, Small, Farmer, and Avouris}}]{lin09}
\bibinfo{author}{\bibfnamefont{Y.-M.} \bibnamefont{Lin}},
  \bibinfo{author}{\bibfnamefont{K.~A.} \bibnamefont{Jenkins}},
  \bibinfo{author}{\bibfnamefont{A.}~\bibnamefont{Valdes-Garcia}},
  \bibinfo{author}{\bibfnamefont{J.~P.} \bibnamefont{Small}},
  \bibinfo{author}{\bibfnamefont{D.~B.} \bibnamefont{Farmer}},
  \bibnamefont{and} \bibinfo{author}{\bibfnamefont{P.}~\bibnamefont{Avouris}},
  \bibinfo{journal}{Nano Lett.} \textbf{\bibinfo{volume}{9}},
  \bibinfo{pages}{422} (\bibinfo{year}{2009}).

\bibitem[{\citenamefont{Neto et~al.}(2009)\citenamefont{Neto, Guinea, Peres,
  Novoselov, and Geim}}]{neto09}
\bibinfo{author}{\bibfnamefont{A.~H.~C.} \bibnamefont{Neto}},
  \bibinfo{author}{\bibfnamefont{F.}~\bibnamefont{Guinea}},
  \bibinfo{author}{\bibfnamefont{N.~M.~R.} \bibnamefont{Peres}},
  \bibinfo{author}{\bibfnamefont{K.~S.} \bibnamefont{Novoselov}},
  \bibnamefont{and} \bibinfo{author}{\bibfnamefont{A.~K.} \bibnamefont{Geim}},
  \bibinfo{journal}{Rev. Mod. Phys.} \textbf{\bibinfo{volume}{81}},
  \bibinfo{pages}{109} (\bibinfo{year}{2009}).

\bibitem[{\citenamefont{Han et~al.}(2007)\citenamefont{Han, {\"O}zyilmaz,
  Zhang, and Kim}}]{han07}
\bibinfo{author}{\bibfnamefont{M.~Y.} \bibnamefont{Han}},
  \bibinfo{author}{\bibfnamefont{B.}~\bibnamefont{{\"O}zyilmaz}},
  \bibinfo{author}{\bibfnamefont{Y.}~\bibnamefont{Zhang}}, \bibnamefont{and}
  \bibinfo{author}{\bibfnamefont{P.}~\bibnamefont{Kim}},
  \bibinfo{journal}{Phys. Rev. Lett.} \textbf{\bibinfo{volume}{98}},
  \bibinfo{pages}{206805} (\bibinfo{year}{2007}).

\bibitem[{\citenamefont{Lemme et~al.}(2007)\citenamefont{Lemme, Echtermeyer,
  Baus, and Kurz}}]{lemme07}
\bibinfo{author}{\bibfnamefont{M.~C.} \bibnamefont{Lemme}},
  \bibinfo{author}{\bibfnamefont{T.~J.} \bibnamefont{Echtermeyer}},
  \bibinfo{author}{\bibfnamefont{M.}~\bibnamefont{Baus}}, \bibnamefont{and}
  \bibinfo{author}{\bibfnamefont{H.}~\bibnamefont{Kurz}},
  \bibinfo{journal}{{IEEE} Electron Device Lett.}
  \textbf{\bibinfo{volume}{28}}, \bibinfo{pages}{282} (\bibinfo{year}{2007}).

\bibitem[{\citenamefont{Avouris et~al.}(2007)\citenamefont{Avouris, Chen, and
  Perebeinos}}]{avouris07}
\bibinfo{author}{\bibfnamefont{P.}~\bibnamefont{Avouris}},
  \bibinfo{author}{\bibfnamefont{Z.}~\bibnamefont{Chen}}, \bibnamefont{and}
  \bibinfo{author}{\bibfnamefont{V.}~\bibnamefont{Perebeinos}},
  \bibinfo{journal}{Nature Nanotech.} \textbf{\bibinfo{volume}{2}},
  \bibinfo{pages}{605} (\bibinfo{year}{2007}).

\bibitem[{\citenamefont{Zhou et~al.}(2007)\citenamefont{Zhou, Gweon, Fedorov,
  First, de~Heer, Lee, Guinea, Neto, and Lanzara}}]{zhou07}
\bibinfo{author}{\bibfnamefont{S.~Y.} \bibnamefont{Zhou}},
  \bibinfo{author}{\bibfnamefont{G.-H.} \bibnamefont{Gweon}},
  \bibinfo{author}{\bibfnamefont{A.~V.} \bibnamefont{Fedorov}},
  \bibinfo{author}{\bibfnamefont{P.~N.} \bibnamefont{First}},
  \bibinfo{author}{\bibfnamefont{W.~A.} \bibnamefont{de~Heer}},
  \bibinfo{author}{\bibfnamefont{D.-H.} \bibnamefont{Lee}},
  \bibinfo{author}{\bibfnamefont{F.}~\bibnamefont{Guinea}},
  \bibinfo{author}{\bibfnamefont{A.~H.~C.} \bibnamefont{Neto}},
  \bibnamefont{and} \bibinfo{author}{\bibfnamefont{A.}~\bibnamefont{Lanzara}},
  \bibinfo{journal}{Nature Mater.} \textbf{\bibinfo{volume}{6}},
  \bibinfo{pages}{770} (\bibinfo{year}{2007}).

\bibitem[{\citenamefont{Reina et~al.}(2009)\citenamefont{Reina, Jia, Ho,
  Nezich, Son, Bulovic, Dresselhaus, and Kong}}]{reina09}
\bibinfo{author}{\bibfnamefont{A.}~\bibnamefont{Reina}},
  \bibinfo{author}{\bibfnamefont{X.}~\bibnamefont{Jia}},
  \bibinfo{author}{\bibfnamefont{J.}~\bibnamefont{Ho}},
  \bibinfo{author}{\bibfnamefont{D.}~\bibnamefont{Nezich}},
  \bibinfo{author}{\bibfnamefont{H.}~\bibnamefont{Son}},
  \bibinfo{author}{\bibfnamefont{V.}~\bibnamefont{Bulovic}},
  \bibinfo{author}{\bibfnamefont{M.~S.} \bibnamefont{Dresselhaus}},
  \bibnamefont{and} \bibinfo{author}{\bibfnamefont{J.}~\bibnamefont{Kong}},
  \bibinfo{journal}{Nano Lett.} \textbf{\bibinfo{volume}{9}},
  \bibinfo{pages}{30} (\bibinfo{year}{2009}).

\bibitem[{\citenamefont{Liang et~al.}(2007)\citenamefont{Liang, Fu, and
  Chou}}]{liangfu07}
\bibinfo{author}{\bibfnamefont{X.}~\bibnamefont{Liang}},
  \bibinfo{author}{\bibfnamefont{Z.}~\bibnamefont{Fu}}, \bibnamefont{and}
  \bibinfo{author}{\bibfnamefont{S.~Y.} \bibnamefont{Chou}},
  \bibinfo{journal}{Nano Lett.} \textbf{\bibinfo{volume}{7}},
  \bibinfo{pages}{3840} (\bibinfo{year}{2007}).

\bibitem[{\citenamefont{Li et~al.}(2008{\natexlab{a}})\citenamefont{Li, Windl,
  and Padture}}]{lidong08}
\bibinfo{author}{\bibfnamefont{D.}~\bibnamefont{Li}},
  \bibinfo{author}{\bibfnamefont{W.}~\bibnamefont{Windl}}, \bibnamefont{and}
  \bibinfo{author}{\bibfnamefont{N.~P.} \bibnamefont{Padture}},
  \bibinfo{journal}{Adv. Mater.} \textbf{\bibinfo{volume}{20}},
  \bibinfo{pages}{1} (\bibinfo{year}{2008}{\natexlab{a}}).

\bibitem[{\citenamefont{Liang et~al.}(2009)\citenamefont{Liang, Chang, Zhang,
  Harteneck, Choo, Olynick, and Cabrini}}]{liang09}
\bibinfo{author}{\bibfnamefont{X.}~\bibnamefont{Liang}},
  \bibinfo{author}{\bibfnamefont{A.~S.~P.} \bibnamefont{Chang}},
  \bibinfo{author}{\bibfnamefont{Y.}~\bibnamefont{Zhang}},
  \bibinfo{author}{\bibfnamefont{B.~D.} \bibnamefont{Harteneck}},
  \bibinfo{author}{\bibfnamefont{H.}~\bibnamefont{Choo}},
  \bibinfo{author}{\bibfnamefont{D.~L.} \bibnamefont{Olynick}},
  \bibnamefont{and} \bibinfo{author}{\bibfnamefont{S.}~\bibnamefont{Cabrini}},
  \bibinfo{journal}{Nano Lett.} \textbf{\bibinfo{volume}{9}},
  \bibinfo{pages}{467} (\bibinfo{year}{2009}).

\bibitem[{\citenamefont{Tung et~al.}(2009)\citenamefont{Tung, Allen, Yang, and
  Kaner}}]{tung08}
\bibinfo{author}{\bibfnamefont{V.~C.} \bibnamefont{Tung}},
  \bibinfo{author}{\bibfnamefont{M.~J.} \bibnamefont{Allen}},
  \bibinfo{author}{\bibfnamefont{Y.}~\bibnamefont{Yang}}, \bibnamefont{and}
  \bibinfo{author}{\bibfnamefont{R.~B.} \bibnamefont{Kaner}},
  \bibinfo{journal}{Nature Nanotech.} \textbf{\bibinfo{volume}{4}},
  \bibinfo{pages}{25} (\bibinfo{year}{2009}).

\bibitem[{\citenamefont{Geim and Novoselov}(2007)}]{geim07}
\bibinfo{author}{\bibfnamefont{A.~K.} \bibnamefont{Geim}} \bibnamefont{and}
  \bibinfo{author}{\bibfnamefont{K.~S.} \bibnamefont{Novoselov}},
  \bibinfo{journal}{Nature Mater.} \textbf{\bibinfo{volume}{6}},
  \bibinfo{pages}{183} (\bibinfo{year}{2007}).

\bibitem[{\citenamefont{Li et~al.}(2008{\natexlab{b}})\citenamefont{Li, Wang,
  Zhang, Lee, and Dai}}]{li08}
\bibinfo{author}{\bibfnamefont{X.}~\bibnamefont{Li}},
  \bibinfo{author}{\bibfnamefont{X.}~\bibnamefont{Wang}},
  \bibinfo{author}{\bibfnamefont{L.}~\bibnamefont{Zhang}},
  \bibinfo{author}{\bibfnamefont{S.}~\bibnamefont{Lee}}, \bibnamefont{and}
  \bibinfo{author}{\bibfnamefont{H.}~\bibnamefont{Dai}},
  \bibinfo{journal}{Science} \textbf{\bibinfo{volume}{319}},
  \bibinfo{pages}{1229} (\bibinfo{year}{2008}{\natexlab{b}}).

\bibitem[{\citenamefont{Sordan et~al.}(2009)\citenamefont{Sordan, Traversi, and
  Russo}}]{sordan09}
\bibinfo{author}{\bibfnamefont{R.}~\bibnamefont{Sordan}},
  \bibinfo{author}{\bibfnamefont{F.}~\bibnamefont{Traversi}}, \bibnamefont{and}
  \bibinfo{author}{\bibfnamefont{V.}~\bibnamefont{Russo}},
  \bibinfo{journal}{Appl. Phys. Lett.} \textbf{\bibinfo{volume}{94}},
  \bibinfo{pages}{073305} (\bibinfo{year}{2009}).

\bibitem[{\citenamefont{Kang and Leblebici}(2002)}]{kang02}
\bibinfo{author}{\bibfnamefont{S.-M.} \bibnamefont{Kang}} \bibnamefont{and}
  \bibinfo{author}{\bibfnamefont{Y.}~\bibnamefont{Leblebici}},
  \emph{\bibinfo{title}{{CMOS} Digital Integrated Circuits Analysis Design}}
  (\bibinfo{publisher}{McGraw-Hill}, \bibinfo{address}{New York},
  \bibinfo{year}{2002}).

\bibitem[{\citenamefont{Ferrari et~al.}(2006)\citenamefont{Ferrari, Meyer,
  Scardaci, Casiraghi, Lazzeri, Mauri, Piscanec, Jiang, Novoselov, Roth
  et~al.}}]{ferrari06}
\bibinfo{author}{\bibfnamefont{A.~C.} \bibnamefont{Ferrari}},
  \bibinfo{author}{\bibfnamefont{J.~C.} \bibnamefont{Meyer}},
  \bibinfo{author}{\bibfnamefont{V.}~\bibnamefont{Scardaci}},
  \bibinfo{author}{\bibfnamefont{C.}~\bibnamefont{Casiraghi}},
  \bibinfo{author}{\bibfnamefont{M.}~\bibnamefont{Lazzeri}},
  \bibinfo{author}{\bibfnamefont{F.}~\bibnamefont{Mauri}},
  \bibinfo{author}{\bibfnamefont{S.}~\bibnamefont{Piscanec}},
  \bibinfo{author}{\bibfnamefont{D.}~\bibnamefont{Jiang}},
  \bibinfo{author}{\bibfnamefont{K.~S.} \bibnamefont{Novoselov}},
  \bibinfo{author}{\bibfnamefont{S.}~\bibnamefont{Roth}}, \bibnamefont{et~al.},
  \bibinfo{journal}{Phys. Rev. Lett.} \textbf{\bibinfo{volume}{97}},
  \bibinfo{pages}{187401} (\bibinfo{year}{2006}).

\bibitem[{\citenamefont{Liu et~al.}(2008)\citenamefont{Liu, Ryu, Tomasik,
  Stolyarova, Jung, Hybertsen, Steigerwald, Brus, and Flynn}}]{liu08}
\bibinfo{author}{\bibfnamefont{L.}~\bibnamefont{Liu}},
  \bibinfo{author}{\bibfnamefont{S.}~\bibnamefont{Ryu}},
  \bibinfo{author}{\bibfnamefont{M.~R.} \bibnamefont{Tomasik}},
  \bibinfo{author}{\bibfnamefont{E.}~\bibnamefont{Stolyarova}},
  \bibinfo{author}{\bibfnamefont{N.}~\bibnamefont{Jung}},
  \bibinfo{author}{\bibfnamefont{M.~S.} \bibnamefont{Hybertsen}},
  \bibinfo{author}{\bibfnamefont{M.~L.} \bibnamefont{Steigerwald}},
  \bibinfo{author}{\bibfnamefont{L.~E.} \bibnamefont{Brus}}, \bibnamefont{and}
  \bibinfo{author}{\bibfnamefont{G.~W.} \bibnamefont{Flynn}},
  \bibinfo{journal}{Nano Lett.} \textbf{\bibinfo{volume}{8}},
  \bibinfo{pages}{1965} (\bibinfo{year}{2008}).

\bibitem[{\citenamefont{Moser et~al.}(2007)\citenamefont{Moser, Barreiro, and
  Bachtold}}]{moser07}
\bibinfo{author}{\bibfnamefont{J.}~\bibnamefont{Moser}},
  \bibinfo{author}{\bibfnamefont{A.}~\bibnamefont{Barreiro}}, \bibnamefont{and}
  \bibinfo{author}{\bibfnamefont{A.}~\bibnamefont{Bachtold}},
  \bibinfo{journal}{Appl. Phys. Lett.} \textbf{\bibinfo{volume}{91}},
  \bibinfo{pages}{163513} (\bibinfo{year}{2007}).

\bibitem[{\citenamefont{Ishigami et~al.}(2009)\citenamefont{Ishigami, Chen,
  Cullen, Fuhrer, and Williams}}]{ishigami07}
\bibinfo{author}{\bibfnamefont{M.}~\bibnamefont{Ishigami}},
  \bibinfo{author}{\bibfnamefont{J.~H.} \bibnamefont{Chen}},
  \bibinfo{author}{\bibfnamefont{W.~G.} \bibnamefont{Cullen}},
  \bibinfo{author}{\bibfnamefont{M.~S.} \bibnamefont{Fuhrer}},
  \bibnamefont{and} \bibinfo{author}{\bibfnamefont{E.~D.}
  \bibnamefont{Williams}}, \bibinfo{journal}{Nano Lett.}
  \textbf{\bibinfo{volume}{7}}, \bibinfo{pages}{1643} (\bibinfo{year}{2009}).

\bibitem[{\citenamefont{Martin et~al.}(2008)\citenamefont{Martin, Akerman,
  Ulbricht, Lohmann, Smet, Klitzing, and Yacoby}}]{martin08}
\bibinfo{author}{\bibfnamefont{J.}~\bibnamefont{Martin}},
  \bibinfo{author}{\bibfnamefont{N.}~\bibnamefont{Akerman}},
  \bibinfo{author}{\bibfnamefont{G.}~\bibnamefont{Ulbricht}},
  \bibinfo{author}{\bibfnamefont{T.}~\bibnamefont{Lohmann}},
  \bibinfo{author}{\bibfnamefont{J.~H.} \bibnamefont{Smet}},
  \bibinfo{author}{\bibfnamefont{K.~V.} \bibnamefont{Klitzing}},
  \bibnamefont{and} \bibinfo{author}{\bibfnamefont{A.}~\bibnamefont{Yacoby}},
  \bibinfo{journal}{Nature Phys.} \textbf{\bibinfo{volume}{4}},
  \bibinfo{pages}{144} (\bibinfo{year}{2008}).

\bibitem[{\citenamefont{Mistry et~al.}(2007)\citenamefont{Mistry, Allen, Auth,
  Beattie, Bergstrom, Bost, Brazier, Buehler, Cappellani, Chau
  et~al.}}]{mistry07}
\bibinfo{author}{\bibfnamefont{K.}~\bibnamefont{Mistry}},
  \bibinfo{author}{\bibfnamefont{C.}~\bibnamefont{Allen}},
  \bibinfo{author}{\bibfnamefont{C.}~\bibnamefont{Auth}},
  \bibinfo{author}{\bibfnamefont{B.}~\bibnamefont{Beattie}},
  \bibinfo{author}{\bibfnamefont{D.}~\bibnamefont{Bergstrom}},
  \bibinfo{author}{\bibfnamefont{M.}~\bibnamefont{Bost}},
  \bibinfo{author}{\bibfnamefont{M.}~\bibnamefont{Brazier}},
  \bibinfo{author}{\bibfnamefont{M.}~\bibnamefont{Buehler}},
  \bibinfo{author}{\bibfnamefont{A.}~\bibnamefont{Cappellani}},
  \bibinfo{author}{\bibfnamefont{R.}~\bibnamefont{Chau}}, \bibnamefont{et~al.},
  \bibinfo{journal}{Tech. Dig. -- Int. Electron Devices Meet.} pp.
  \bibinfo{pages}{247--250} (\bibinfo{year}{2007}).

\end{thebibliography}
\end{document}